\documentclass[preprint,preprintnumbers,amsmath,amssymb]{revtex4-1}

\usepackage[dvipsnames]{xcolor}
\usepackage{color}
\usepackage{slashed}
\usepackage{amsmath}
\usepackage{amsthm}
\usepackage{subfigure}
\usepackage[utf8]{inputenc}
\usepackage{graphicx}
\usepackage{epsfig}
\usepackage{amssymb}
\usepackage{dcolumn}
\usepackage{bm}

\newcommand{\ba}{\begin{array}}
\newcommand{\ea}{\end{array}}
\def\br{\begin{eqnarray}}
\def\er{\end{eqnarray}}
\def\be{\begin{equation}}
\def\ee{\end{equation}}
\usepackage{hyperref}

\def\({\left(}
\def\){\right)}

\def\<{\left\langle}
\def\>{\right\rangle}

\begin{document}

\title{Schwinger-Dyson equation boundary conditions induced by ETC radiative corrections }

\date{\today}

\begin{abstract}

The technicolor (TC)  Schwinger-Dyson equations (SDE) should include radiative corrections induced by extended technicolor (ETC) interactions when TC is embedded into a larger theory including also QCD. These radiative corrections couple the different strongly interacting Dyson equations. We discuss how the boundary conditions of the coupled SDE system are modified by these corrections, and verify that the ultraviolet  behavior of the self-energies are described by a function that decreases logarithmically with momentum. 

\end{abstract}

\pacs{12.60.Cn, 12.60.Rc, 11.30.Na}

\author{A. Doff}
\email{agomes@utfpr.edu.br}

\affiliation{Universidade Tecnol\'ogica Federal do Paran\'a - UTFPR - DAFIS
Av Monteiro Lobato Km 04, 84016-210, Ponta Grossa, PR, Brazil}

\author{A. A. Natale} 
\email{natale@ift.unesp.br}

\affiliation{Instituto de F{\'i}sica Te\'orica, UNESP, Rua Dr. Bento T. Ferraz, 271, Bloco II, 01140-070, S\~ao Paulo, SP, Brazil}



\maketitle

\section{Introduction}

\par The chiral and gauge symmetry breaking in quantum field theories can be promoted by fundamental scalar bosons through the Higgs boson mechanism. If this particle is a composite or an elementary scalar boson is still an open question. Many models have considered the possibility of a light composite Higgs based on effective Higgs potentials as reviewed in Ref.\cite{h1}.  Nambu and Jona-Lasinio proposed one of the first field theoretical models based on the ideas of superconductivity, where all the most important aspects of chiral symmetry breaking and mass generation, as known nowadays, were explored at length \cite{nl}. The model of Ref.\cite{nl}
contains only fermions possessing invariance under chiral symmetry, although this invariance is not respected by the vacuum of the theory
and the fermions acquire a dynamically generated mass. As a consequence of the chiral symmetry breaking by the vacuum the
analysis of the Bethe-Salpeter equation (BSE) shows the presence of Goldstone bosons. These bosons, when the theory is assumed to be
the effective theory of strongly interacting hadrons, are associated to the pions. Besides these aspects Nambu and Jona-Lasinio also verified that the theory presents
a scalar bound state (the sigma meson), which plays the role of the Higgs boson in their strong interaction model.   

In Quantum Chromodynamics (QCD) the same mechanism is observed, where the quarks acquire a dynamically generated mass ($\mu$).
This dynamical mass is usually expected to appear as a solution of the SDE for the fermion propagator when the
coupling constant is above a certain critical value. The same condition that leads to chiral symmetry breaking is also responsible
to generate a bound-state massless pion, and a scalar p-wave state of the BSE, indicating the
presence of a scalar state with mass $m_\sigma = 2 \mu$. This scalar meson is the elusive sigma meson \cite{pl1,pl2,pl3}, that is assumed
to be the Higgs boson of QCD. This scenario is the accomplishment of Nambu and Jona-Lasinio proposal in the context of renormalizable gauge theories.

The possibility of spontaneous gauge and chiral symmetry breaking promoted by a composite scalar boson in the context of the 
Standard Model (SM) was formulated in the seventies by Weinberg \cite{we} and Susskind \cite{su}. The most popular 
version of these models was dubbed as technicolor (TC), where new fermions (or technifermions) condensate and may be 
responsible for the chiral and SM gauge symmetry breaking \cite{Rev1,Rev2,Rev3}.
However the phenomenology of these models depend crucially on these new fermions (or technifermions) self-energy. In the early models this self-energy was considered to be given by the result \cite{Lane,ope}
\be
\Sigma_{TC} (p^2) \propto \frac{\left\langle {\bar{T}_f}T_f\right\rangle_{\mu}}{p^2} \,\, ,
\label{eq0}
\ee
where $\left\langle {\bar{T}_f}T_f\right\rangle_\mu \sim \mu^3$ is the TC condensate and $\mu$ is the characteristic TC dynamical mass scale, which is of order of a few hundred GeV, i.e. the order of the SM vacuum expectation value. Unfortunately early technicolor models suffered from problems like flavor changing neutral currents (FCNC) and contributions to the electroweak corrections not compatible with the experimental data \cite{Rev1,Rev2,Rev3}. These problems occur when new extended technicolor interactions (ETC) are introduced in order
to provide masses to the standard quarks. Eq.(\ref{eq0}) leads to quark masses that vary with the ETC mass scale ($M_E$) as $1/M_E^2$.

\par  A possible way out of this dilemma was proposed by Holdom\cite{holdom} many years ago, remembering that the self-energy behaves as
\be 
\Sigma_{TC} (p^2)\approx \frac{\left\langle {\bar{T}_f}T_f\right\rangle_{\mu}}{p^2} \left(\frac{p}{\mu}\right)^{\gamma_m} \,\, 
\label{eqa}
\ee
where $\gamma_m$ the mass anomalous dimension associated to the fermionic condensate. As can be verified from Eq.(\ref{eqa}) a large anomalous dimension leads to a hard asymptotic self-energy (or quasi-conformal technicolor theories) and  this may solve the many problems of the SM symmetry breaking promoted by composite bosons\cite{walk2,walk3,walk4,walk5,walk6,walk7,walk8,walk9}. Quark masses will be less dependent
on the ETC interactions in the case of a hard TC self-energy, leading to a less problematic phenomenology.

There are different ways of obtaining a large $\gamma_m$ value in Eq.(\ref{eqa}), in what is known as extreme walking (or quasi-conformal) 
TC  theories. (i) It is possible to obtain an almost conformal TC theory when the fermions are in the fundamental representation introducing a large number of TC fermions ($n_{TF}$), leading to an almost zero $\beta$ function and flat asymptotic coupling constant. The cost of such procedure may be a large S parameter \cite{peskin1,peskin2}, such behavior can also be obtained  when the fermions are in larger
representations other than the fundamental one \cite{sannino1,sannino2,sannino3}; or (ii) by inclusion of four-fermion 
interactions\cite{yama1, yama2, mira2, yama3, mira3, yama4}. 

Most of these studies were performed looking at SDE solutions of the technifermion propagator. In particular, after a work by 
Takeuchi \cite{tak}, it became clear that the technifermion self-energy may vary between the behavior of Eq.(\ref{eq0}) and 
the extreme behavior, that in the past was called irregular solution \cite{Lane}, that is giving by
\be 
\Sigma_{TC} (p^2)\approx \mu\left[1 + bg^2 (\mu^2) \ln\left(p^2/\mu^2 \right) \right]^{-\delta} \,, 
\label{eqb}
\ee 
where  in Eq.(\ref{eqb}) $g$ is the TC running coupling constant, $b$ is the coefficient of $g^3$ term in the renormalization group $\beta (g)$ function, $\delta= \frac{3c}{16\pi^2 b}$ , and  $c$ is the quadratic Casimir operator given by 
$$
c = \frac{1}{2}\left[C_{2}(R_{1}) +  C_{2}(R_{2}) - C_{2}(R_{3})\right] \,\,\, ,
$$
\noindent and $C_{2}(R_{i})$,  are the Casimir operators for fermions in the representations  $R_{1}$ and $R_{2}$ that form a composite boson in the representation $R_{3}$. The behavior of Eq.(\ref{eqb}) happens when the theory is totally dominated by a four-fermion interaction,
like in the Nambu-Jona-Lasinio model, and it is quite interesting because it may lead
to a composite TC scalar boson much lighter than the TC characteristic scale \cite{us1,us2,us4,twoscale,us3}. Eq.(\ref{eqb}) leads to quark masses that vary with the ETC mass scale as $[\ln (\mu^2/M_E^2)]^{-\delta}$.

It is not surprising that the introduction of a four-fermion interaction may change the ultraviolet SDE behavior. As observed
by Cohen and Georgi \cite{cg} much of the information about chiral symmetry breaking resides into the boundary conditions, and
the introduction of new interactions change these conditions. Recently we discussed how the boundary conditions of the  anharmonic oscillator representation of the 
SDE for $SU(N)$ gauge theories are directly related to, and may change, the mass anomalous 
dimensions \cite{us5}. Motivated by this we studied how the introduction of radiative corrections into the SDE
may change the self-energy solutions \cite{ardn}, and verified that when TC is embedded into a larger theory including also QCD,  radiative corrections couple the different strongly interacting Dyson equations (TC and QCD) and change completely the ultraviolet behavior of the gap equation solution. The work of Ref.\cite{ardn} was performed numerically and we just commented, without presenting the details of the 
calculation, that the effect of the radiative corrections in the coupled equations was similar to a change in the anomalous mass dimension of the theory. The purpose of this work is to show in detail how the coupled TC and QCD have their boundary conditions changed by
the ETC radiative corrections, in such a way that the self-energies ultraviolet behavior turn out to be of the form that we may
call extreme walking or irregular one, i.e. the behavior of Eq.(\ref{eqb}), what may indicate a new way to build TC models as described in Ref.\cite{ardn}.

This  work is organized as follows, in section II we present the TC and QCD coupled SDE system discussed in Ref.\cite{ardn}, we transform the integral SDE equations
into a pair of differential equations, and considering some approximate analytical expressions we recover the numerical result of Ref.\cite{ardn}, where the quark mass is totally dominated by the irregular solution given by Eq.(\ref{eqb}), and in section III we verify how the boundary conditions of the gap equations are
changed and are, consequently, responsible by the different asymptotic self-energies behavior. In Section IV we draw our conclusions. 

\section{TC and QCD coupled SDE system by ETC interactions}

\par  In Ref.\cite{ardn} we  discussed a coupled SDE system where two strongly interacting theories, TC and QCD, are interconnected
by corrections due to ETC and other interactions. These SDE are displayed in Fig.(1). These diagrams appear naturally when QCD and
TC are embedded into a larger gauge group, like, for instance, the $SU(5)_S$ Farhi-Susskind model \cite{far} which plays the
role of the ETC group. These gap equations may also contain electroweak corrections and, as we are not specifying a model, other possible interactions, which may contribute to the last diagram on the right-hand side of Fig.(1). 

In the work of Ref.\cite{ardn} the SDE equations were solved numerically in a quite simplified approximation, using bare vertices and gauge boson propagators, and verified that the ultraviolet self-energy behaviors of quarks and techniquarks are changed from a soft to a hard
behavior, i.e. from a fast to a slow decrease of the self-energy with the momentum. Here we will focus in a detailed analytical approach in order to show that the change in the asymptotic behavior of the self-energies, is a direct consequence of a change of the SDE boundary 
conditions due to the ETC radiative corrections. In order to do so most of this section is devoted to a detailed discussion 
of the coupled SDE and to write them in a differential and dimensionless form in order to expose their dependence on the boundary
conditions.

\begin{figure}[t]
\centering
\includegraphics[width=1\columnwidth]{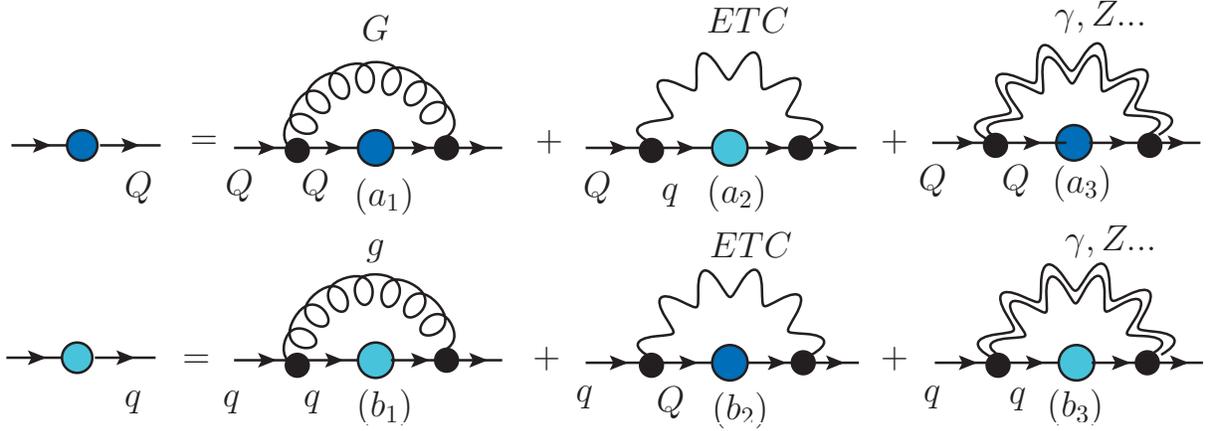}
\caption[dummy0]{ The Schwinger-Dyson equations\cite{ardn} for QCD(q$\equiv$quark) and TC(Q$\equiv$technifermion) including  ETC and electroweak or other corrections($\gamma,Z..$). In the above figure $G \,(g)$ indicate technigluons (gluons) propagators. }
\label{fig1}
\end{figure}

The diagrams denoted by $a_i$ and $b_i$ with $i=1$ in Fig.(1) are respectively the known SDE for techniquarks and quarks. These equations
become coupled through the ETC interactions as indicated by the diagrams $a_2$ and $b_2$. We shall not discuss the effect of 
diagrams $a_3$ and $b_3$, which were briefly discussed in Ref.\cite{ardn}. The TC SDE (diagram $a_1$), whose self-energy, coupling constant
 and respective Casimir operator will be indicated by the index $(1)$, receives a correction (diagram $a_2$) due to the quarks self-energy indicated by the index $(2)$ with charge $C_E\alpha_E$ and gauge boson mass $M_E$ related to the ETC group, leading to the following equation\cite{ardn}  
\br 
&&\hspace*{-0.5cm}\Sigma_1(p)=\frac{3C_1\alpha_1}{2\pi^2}\int dk^2 dA \frac{k^2\Sigma_1(k^2)}{(p-k)^2[k^2+\Sigma^2_1(k^2)]} + \nonumber \\
&&\hspace*{0.1cm}  \frac{3C_E\alpha_E}{2\pi^2}\int dk^2 dA \frac{k^2\Sigma_{2}(k^2)}{k^2+\Sigma^2_{2}(k^2)}\frac{1}{(p-k)^2 + M^2_E}  ,
\label{eq1}
\er 
\noindent whereas for the quarks self-energy we have a similar equation just changing the index  $1 \leftrightarrow 2$ 
\br 
&&\hspace*{-0.2cm}\Sigma_2(p)=\frac{3C_2\alpha_2}{2\pi^2}\int dk^2 dA \frac{k^2\Sigma_2(k^2)}{(p-k)^2[k^2+\Sigma^2_2(k^2)]}+\nonumber \\
&&\hspace*{0.2cm} \frac{3C_E\alpha_E}{2\pi^2} \int dk^2 dA \frac{k^2\Sigma_1(k^2)}{[(p-k)^2+M_E^2]\cdot[k^2+\Sigma^2_1 (k^2)]}  ,
\label{eq2}
\er
where $dA=d\theta \sin^2{\theta}$.

We can easily identify the second term in right-hand side of Eq.(\ref{eq2}) as the usual quark mass obtained through TC interaction. With the
appropriate QCD values for $C_2 \alpha_2$ and ETC values for $C_E \alpha_E$ and $M_E$ we obtain a solution that is the sum of the dynamical
quark mass with its effective ``bare mass". Also Eq.(\ref{eq1}) provides the dynamical techniquark mass with a very tiny mass generated
by the QCD correction. 

The above equations were solved numerically in Ref.\cite{ardn}. Here they will be transformed into a coupled system of differential
equations, therefore we will need to make a few simplifications and the first one is to perform the angular integration
using the angle approximation\cite{craig},
transforming the following terms as
\br 
\frac{1}{(p - k)^2 + M^2} = \frac{\pi}{2}\left\{\frac{\theta(p - k)}{p^2 + M^2} + \frac{\theta(k - p)}{k^2 + M^2}\right\} ,
\label{eqang} 
\er 
\noindent where in the sequence we may take $M = M_E$ or $M = 0 $. 

Continuing with the notation $(1\equiv TC)$ and $ (2\equiv QCD) $ 
we obtain the following form for the system of coupled integral equations
\br  
&&\hspace*{-0.2cm}f_1(x)= \sigma_1(x)I^a_1(x) + \theta_1I^b_1(x) + \zeta_{12}(x)I^a_2(x) + \eta_{12}I^b_{E2}(x)  \nonumber \\ \nonumber \\
&&\hspace*{-0.2cm}f_2(x)= \sigma_2(x)I^a_2(x) + \theta_2I^b_2(x) + \zeta_{21}(x)I^a_1(x) + \eta_{21}I^b_{E1}(x).  \nonumber \\
\label{A1}
\er
\noindent  To arrive at the last expression we introduced the following set of new variables and auxiliary functions
\br 
&& p^2 = M^2_ix \,\,\,,\,\,\, \Sigma_i(p^2) = M_i f_i(x)\,\,\,,\,\,\, \omega_{i} =\frac{m^2_i}{M^2_i}\nonumber \\
&& k^2 = M^2_iy \,\,\,,\,\,\,\Sigma_i(k^2) = M_i f_i(y)\,\,\,,\,\,\,  \beta_{i} =\frac{M^2}{M^2_i}\nonumber \\
&&\sigma_i(x) = \frac{\theta_i}{x + \omega_i}\,\,\,,\,\,\,\theta_i =\frac{3C_i\alpha_i}{4\pi}\,\,\,,\,\,\,\theta_E =\frac{3C_E\alpha_E}{4\pi} \nonumber \\
&& I^a_i(x) = \int^{x}_{0}dy\frac{yf_i(y)}{y+f^2_i(y)}\,\,\,,\,\,\,g_i(x) = \frac{xf_i(x)}{x+f^2_i(x)}  \nonumber \\ 
&& I^b_i(x) = \int^{\frac{\Lambda^2}{M^2_i}}_{x} dy\frac{yf_i(y)}{y+f^2_ i(y)}\frac{1}{y +\omega_i} \nonumber \\
&& I^b_{Ei}(x) = \int^{\frac{\Lambda^2}{M^2_i}}_{x} dy\frac{yf_i(y)}{y+f^2_ i(y)}\frac{1}{y +\beta_i} \nonumber \\
&& \zeta_{ij}(x) = \frac{\theta_E}{x+\beta_i}\left(\frac{M_j}{M_i}\right)^3\,\,\,,\,\,\,\eta_{ij} = \theta_E\left(\frac{M_j}{M_i}\right).
\label{A2}
\er 
\noindent In the above expressions $M=M_{E}$, $i,j=1(2)$ denote the contributions of TC(QCD) to the coupled gap equation. Note that 
$M_1 = \mu_{{}_{1}}$,  and  $M_2 = \mu_{{}_{2}}$, where $\mu_i$  correspond respectively to the dynamical TC and QCD fermionic mass scales,
$m_i$ represents technigluons(or gluons) dynamical mass scale \cite{mg1,mg2,mg3,mg4}, which were not considered in Ref.\cite{ardn}.

In order to transform the coupled system of integral equations described by Eq.(\ref{A1}) into a system of coupled  differential equations for $f_i(x)$ we also introduce  new
functions $\delta_1(x)$ and $\delta_2(x)$ , where 
\br 
&&\delta_1(x) =\frac{\zeta'_{12}(x) \left(f'_2(x) + \frac{\eta_{21}g_1(x)}{x+\beta_1} - \zeta_{21}(x)g_1(x)\right)}{\zeta'_{12}(x)\zeta'_{21}(x) - \sigma'_2(x)\sigma'_1(x)} \nonumber\\
&& - \frac{\sigma'_2(x)\left(f'_1(x) + \frac{\eta_{12}g_2(x)}{x+\beta_2} - \zeta_{12}(x)g_2(x)\right)}{\zeta'_{12}(x)\zeta'_{21}(x) - \sigma'_2(x)\sigma'_1(x)} 
\label{A3a}
\er 
\br 
&&\delta_2(x)=\frac{ \zeta'_{21}(x) \left(f'_1(x) + \frac{\eta_{12}g_2(x)}{x+\beta_2} - \zeta_{12}(x)g_2(x)\right)}{\zeta'_{12}(x)\zeta'_{21}(x) - \sigma'_2(x)\sigma'_1(x)} \nonumber\\
&& -\frac{\sigma'_1(x)\left(f'_2(x) + \frac{\eta_{21}g_1(x)}{x+\beta_1} - \zeta_{21}(x)g_1(x)\right)}{\zeta'_{12}(x)\zeta'_{21}(x) - \sigma'_2(x)\sigma'_1(x)},\nonumber \\
\label{A3b}
\er
\noindent in such a way that now we can write
\br 
&& f''_1(x) + \frac{\eta_{12}g'_2(x)}{x +\beta_2} + \frac{\zeta_{12}(x)g_2(x)}{x+\beta_1} = \frac{\eta_{12}g_2(x)}{(x+\beta_2)^2} \nonumber \\
&&\hspace*{0.97cm} + \,\zeta_{12}(x)g'_2(x) + \sigma''_1(x)\delta_1(x) + \sigma'_1(x)\delta'_1(x)  \nonumber \\
&&\hspace*{0.97cm} + \frac{{}^{}}{{}^{}}\zeta''_{12}(x)\delta_2(x) + \zeta'_{12}(x)\delta'_2(x)  \nonumber \\
\label{A41}\\ \nonumber \\
&&f''_2(x) + \frac{\eta_{21}g'_1(x)}{x +\beta_1} + \frac{\zeta_{21}(x)g_1(x)}{x+\beta_2} = \frac{\eta_{21}g_1(x)}{(x+\beta_1)^2} \nonumber \\
&&\hspace*{0.97cm} + \, \zeta_{21}(x)g'_1(x) + \sigma''_2(x)\delta_2(x) + \sigma'_2(x)\delta'_2(x)  \nonumber \\
&&\hspace*{0.97cm} + \frac{{}^{}}{{}^{}}\zeta''_{21}(x)\delta_1(x) + \zeta'_{21}(x)\delta'_1(x). \nonumber \\
\label{A42}
\er 

It is an exercise the verification that Eqs.(\ref{A41}) and (\ref{A42}) can be solved by a linear combination of two solutions. 
One that is called regular, where the self-energies behave at large momenta as $1/x$, and another one, called irregular, where the
self-energies decrease as $[\ln x ]^{-\epsilon}$,
where $\epsilon $ is a function of the quantities $b$ and $c$. Only when the boundary conditions are applied to these
equations the actual self-energy ultraviolet behavior is selected. This is the central point of the work of Ref.\cite{ardn}
and the one that we present here: The ETC radiative corrections cause the selection of the irregular solution and not the
one behaving as $1/x$! 

How the ETC corrections will change the SDE boundary conditions and the solution behavior will be
shown in the next section. However, in a very naive approximation for Eq.(\ref{eq2}) we can show in the sequence that 
quark masses vary logarithmically with the ETC mass scale (i.e. $M_E$), which is a consequence of a 
TC self-energy with a logarithmic ultraviolet (UV) behavior \cite{ardn}. As can be seen from Eq.(\ref{eq2}) the full quark mass ($m_q$) is a sum of the 
dynamical mass generated within QCD, with the one generated through TC mediated by the ETC interaction. Using Eq.(\ref{eqang}) and
approximating $m_q$ by $\Sigma_2(0)$ we can simplify Eq.(\ref{eq2}) and obtain 
\br 
&& m_q \approx \Sigma_2 (0) \approx \frac{3C_2\alpha_2}{4\pi}\int_{0}^\Lambda dk^2\frac{\Sigma_2(k)}{k^2 +\Sigma^2_2(k)} \nonumber \\
&&\hspace*{0.6cm}+\frac{3C_E\alpha_E}{4\pi}\int_{0}^\Lambda dk^2\frac{\Sigma_1(k)}{k^2+\Sigma^2_1(k)}\frac{1}{k^2 + M^2_E} \nonumber \\
&&\hspace*{0.6cm}+\frac{3C_2\alpha_2}{4\pi}\frac{1}{M^2_E}\int_{0}^\Lambda dk^2\frac{\Sigma_2(k)}{k^2 +\Sigma^2_2(k)} \nonumber \\
&&\hspace*{0.6cm}+\frac{3C_E\alpha_E}{4\pi}\frac{1}{2M^2_E}\int_{0}^\Lambda dk^2\frac{\Sigma_1(k)}{k^2+\Sigma^2_1(k)} 
\label{eqmq}
\er
This is an oversimplified equation compared to Eq.(\ref{eq2}) and gives $m_q$ as a function of $M_E$. We can now input 
the solutions of Eqs.(\ref{A41}) and (\ref{A42}) into Eq.(\ref{eqmq}) and solve it up to the convergence. The convergence is obtained
only with the solution that has a logarithmic UV behavior. 

Even within the simplified approach shown above it is possible to compare the calculation of Eq.(\ref{eqmq}) with the
full numerical result obtained in Ref.\cite{ardn}. Therefore, assuming $\Lambda = 500$TeV,  
$M_1 = \mu_{1}= 1$TeV, $M_2 =\mu_{2} = 0.3$GeV, $C_1 \alpha_1=6$ and $C_2 \alpha_2=1.16$ values such that isolated techniquark and quark masses were equal respectively to $\mu_1$ and $\mu_2$ as in Ref.\cite{ardn}, and $C_E \alpha_E = 0.032$, we plot $m_q$ as a function
of $M_E$ in Fig.(2).

\begin{figure}[t]
\centering
\includegraphics[width=1.3\columnwidth]{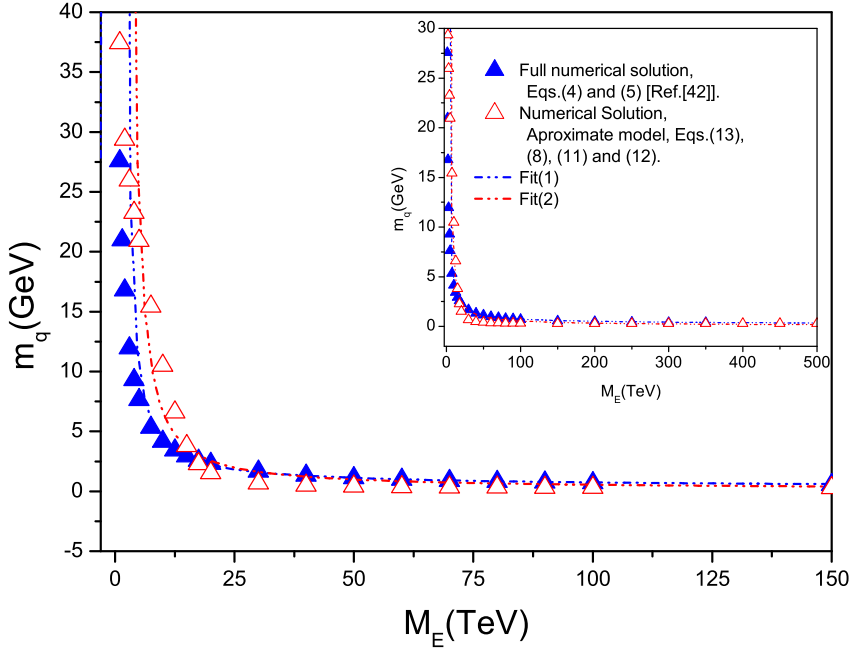}
\vspace*{-3cm}
\caption[dummy0]{Behavior of $m_q$ as a function of $M_E$ obtained from Eq.(\ref{eqmq}) using the solutions of Eqs.(\ref{A42}) and 
(\ref{A41}) (red curve($\textcolor{red}{\Delta}$)). The  blue curve($\textcolor{blue}{\blacktriangle}$) corresponds to the  result obtained in \cite{ardn}. The dot-dash lines  represents  the  Fit(1)(blue dot-dash line) and Fit(2) (Red dot-dash line) , the  parameters are described in the text. }
\label{fig2}
\end{figure}

\par The  blue curve($\textcolor{blue}{\blacktriangle}$) corresponds to the  result obtained in \cite{ardn}, while the red curve($\textcolor{red}{\Delta}$)   is the numerical one obtained with Eqs.(\ref{A41}) , (\ref{A42}) and  (\ref{eqmq}). The dot-dash lines  represent  the fit
\be
m_q^{fit} \propto a_1 [ln(M^2_E/M^2_{2})]^{-a_2} \, ,
\label{eqx5}
\ee
where for the Fit(1) (blue dot-dash) we obtain , $a_1= 203.92$GeV and $a_2= 2.53$ with $R^2=0.99$  and for 
Fit(2) (Red dot-dash) , $a_1= 1620$GeV and $a_2=3.6$ with $R^2=0.93$. The behavior exhibited by the curves in Red depicted in 
Fig.(\ref{fig2}) are very similar
to the one obtained in the Ref.\cite{ardn}. The small discrepancy between these different results can be credited to the
angle approximation and the simple approximation made here to calculate $m_q$. However, the main point is that in TC models as proposed here,
the dependence of quark masses on the ETC mass scale is definitively logarithmic and not a power law, which is a consequence
of quark masses computed with a techniquark self-energy as the one shown in Eq.(\ref{eqb}). 
In the next section we show that both self-energies (techniquarks and quarks)
have the same asymptotic behavior induced by the ETC interaction.

\section{UV boundary conditions induced by ETC radiative corrections}

\par In order to determine the boundary conditions of our SDE coupled system we can differentiate Eqs.(\ref{A1}) obtaining  
\br 
&& f'_1(x)  = \sigma'_1(x)I^a_1(x) + \zeta'_{12}(x)I^a_2(x)
\nonumber \\ 
&& f'_2(x)  = \sigma'_2(x)I^a_2(x) + \zeta'_{21}(x)I^a_1(x),\nonumber \\
\label{e1}
\er
\noindent from these expressions and reconsidering Eq.(\ref{A1}) in the asymptotic UV limit, such that 
$(\frac{\Lambda^2}{M^2_i} = \frac{M^2_E}{M^2_i} >> 1)$ we end up the following set of coupled equations
\br 
&& f_1(x) + xf'_1(x) = \zeta_{12}(x)I^a_2(x)\nonumber \\
&& f_2(x) + xf'_2(x) = \zeta_{21}(x)I^a_1(x)   
\label{e2}
\er
\noindent which correspond to the UV boundary conditions that should be satisfied by Eqs.(\ref{A41}) and (\ref{A42}). Reconsidering  the definitions of variables and auxiliary functions described  in Eq.(\ref{A2}) the above expressions above lead to 
\br 
&& \left[ \Sigma_1(p^2) + p^2\Sigma'_1(p^2)\right]_{p^2\rightarrow \infty} = m_2 ,
\nonumber \\
&& \left[ \Sigma_2(p^2) + p^2\Sigma'_2(p^2)\right]_{p^2\rightarrow \infty} = m_1 ,\nonumber \\
\label{e2l}
\er
\noindent where 
\be
m_{1(2)}=\frac{3C_E\alpha_E}{4\pi}\int dk^2\frac{k^2\Sigma_{1(2)}(k^2)}{k^2+\Sigma^2_{1(2)}(k^2)}\frac{1}{k^2 + M^2_E} .
\label{e4}
\ee

It is easy to recognize that Eqs.(\ref{e2l}) when $\alpha_E =0$, i.e. when the equations are decoupled, we have
\br 
&& \left[ \Sigma_1(p^2) + p^2\Sigma'_1(p^2)\right]_{p^2\rightarrow \infty} = 0,
\nonumber \\
&& \left[ \Sigma_2(p^2) + p^2\Sigma'_2(p^2)\right]_{p^2\rightarrow \infty} = 0 ,\nonumber \\
\label{e2l2}
\er
and it has long been known that the asymptotic behavior of $\Sigma_1(p^2)$ (or $\Sigma_2(p^2)$)
is described by\cite{Lane} 

\be
\Sigma_{1(2)}(p^2 \rightarrow\infty )\propto \frac{\mu^3_{1(2)}}{p^2}
\label{e5}
\ee 
\noindent which lead to quark masses $m_q\approx \Sigma_2 (0)$ of order
\be
m_q \approx C_E\alpha_E\frac{M^3_1}{M_E^2},
\label{e51}
\ee
which is at the origin of all known problems of TC models \cite{Rev1}.

When we turn on the ETC interaction, i.e. $\alpha_E \neq 0$, the equations (\ref{e2l}) differ only by their
infrared (IR) boundary conditions, which are usually set as $\Sigma_1 (0)=M_1$ and $\Sigma_2 (0)=M_2$. If, apart only from
different numerical scales, we set $\Sigma_1 (0)=\Sigma_2 (0)\equiv M $ eqs.(\ref{e2l}) will give
\br 
&& \left[ \Sigma_1(p^2) + p^2\Sigma'_1(p^2)\right]_{p^2\rightarrow \infty} =  M(p^2\rightarrow \infty) ,
\nonumber \\
&& \left[ \Sigma_2(p^2) + p^2\Sigma'_2(p^2)\right]_{p^2\rightarrow \infty} =  M(p^2\rightarrow \infty) ,\nonumber \\
\label{e2ll}
\er
where $ M(p^2\rightarrow \infty) \propto M[\ln p^2/M^2 ]^{-\epsilon}$ and there is no doubt that $\Sigma_1(p^2)$ and $\Sigma_2(p^2)$ have exactly the same IR and UV behavior,
and when $M_1\neq M_2$ they just differ numerically in the IR but have the same functional expression. This
is a confirmation of the statement in Ref.\cite{ardn} that in the scenario proposed there the TC and QCD
self-energies have exactly the same UV asymptotic behavior.

If we do not assume the same IR conditions for both self-energies we can again verify that $\Sigma_1(p^2)$ and $\Sigma_2(p^2)$ 
have formally the same UV behavior. Initially we can observe that in the deep Euclidean region we have
\be
f_i (x \rightarrow\infty ) \approx \frac{\theta_i}{x} \delta_i (x),
\label{e71}
\ee
but in the same limit and after some algebra we can see that $\delta_1 (x)$ and $\delta_2 (x)$ have also the
same expression apart from a constant, therefore
\be
\left. f_1 (x)\right|_{x\rightarrow\infty} \propto \left. f_2 (x)\right|_{x\rightarrow\infty},
\label{e72}
\ee
indicating that both self-energies decrease equally in the UV region.

The main difference in the UV boundary conditions in the decoupled and coupled SDE, assuming that $\Sigma_1 (p^2)$
and $\Sigma_2 (p^2)$ have the same formal expression and both can be substituted by an expression $\Sigma (p^2)$, is an 
effective mass term
\be
m=\frac{3C_E\alpha_E}{4\pi}\int dk^2\frac{k^2\Sigma (k^2)}{k^2+\Sigma^2 (k^2)}\frac{1}{k^2 + M^2_E} ,
\label{e8}
\ee
and this term is going to act like a ``bare" mass, whose effect is to generate a logarithmically decreasing self-energy for quarks and
techniquarks.

\section{Conclusions}

In the Ref.\cite{ardn} we have given evidences that radiative corrections to TC(QCD) change the UV technifermion(quark) self-energy behavior. 
This happens when TC and QCD are embedded into an unified theory as in the Farhi-Susskind model (or an ETC model). In this work we verify that in these cases the radiative corrections that couple the different strongly interacting Dyson equations induce new boundary conditions for the gap equations and change the UV behavior compared to the isolated equations.

We transformed the coupled TC and QCD coupled equations into a pair of differential equations. These two equations may have as a solution
a linear combination of the known regular and irregular self-energies. When these solutions are applied to a quite simplified mass
equation (Eq.(\ref{eqmq})), derived from the original gap equation, we verified that quark masses vary logarithmically with the ETC scale,
what is a consequence of a TC self-energy that also decreases logarithmically with the momentum. This is a simple confirmation of
the more complete numerical calculation of Ref.\cite{ardn}. The simple quark mass (Eq.(\ref{eqmq})) acts as a constraint for
the differential equations solution, appearing as an effective mass boundary condition.

In Section III we discussed how the ETC interaction induce a change into the boundary conditions, and this change is equivalent
to the addition of an effective bare mass to the gap equation, which leads naturally to a self-energy logarithmically decreasing 
with the momentum. Moreover, as stated in Ref.\cite{ardn}, we also discussed that both self-energies have the same formal 
expression and, along with the discussion of that same reference, may lead to a new way to build TC models.

\section*{Acknowledgments}

This research was  partially supported by the Conselho Nacional de Desenvolvimento Cient\'{\i}fico e Tecnol\'ogico (CNPq) under
the grants 302663/2016-9 (AD) and 302884/2014-9 (AAN).


\begin {thebibliography}{99}

\bibitem{h1} B. Bellazzini, C. Cs{\'{a}}ki and J. Serra, Eur. Phys. J. C {\bf 74}, 2766 (2014).

\bibitem{nl} Y. Nambu and G. Jona-Lasinio, {\it Phys. Rev.} {\bf 122}, 345  (1961).

\bibitem{ds} R. Delbourgo and M. D. Scadron, {\it  Phys. Rev. Lett.} {\bf 48}, 379 (1982).

\bibitem{pl1} N. A. Tornqvist and M. Roos, {\it Phys. Rev. Lett.} {\bf 76}, 1575 (1996).

\bibitem{pl2} N. A. Tornqvist and A. D. Polosa, {\it  Nucl. Phys. A} {\bf 692}, 259 (2001).

\bibitem{pl3} N. A. Tornqvist and A. D. Polosa, {\it Frascati Phys. Ser. }{\bf 20}, 385 (2000).

\bibitem{we} S. Weinberg, Phys. Rev. D {\bf 19} 1277 (1979).

\bibitem{su} L. Susskind, Phys. Rev. D {\bf 20}, 2619 (1979).

\bibitem{Rev1}  C. T. Hill and E. H. Simmons, Phys. Rept. {\bf 381}, 235 (2003) [Erratum-ibid. {\bf 390}, 553 (2004)].

\bibitem{Rev2}  F. Sannino,  hep-ph/0911.0931, {\it Lectures presented at the 49th Cracow School of Theoretical Physics. Conformal Dynamics for TeV Physics and Cosmology,  Cracow, Nov , 2009}; Acta Phys. Polon.. B {\bf 40}, 3533 (2009); Int. J. Mod. Phys. A {\bf 20}, 6133 (2005). 

\bibitem{Rev3}  K. Lane, {\it Technicolor 2000 }, Lectures at the LNF Spring School in Nuclear, Subnuclear and Astroparticle Physics, Frascati (Rome), Italy, May 15-20, 2000. 

\bibitem{Lane} K. Lane, {\it  Phys. Rev. D} \textbf{10}, 2605 (1974).

\bibitem{ope} H. D. Politzer, Nucl. Phys. B {\bf 117}, 397 (1976).

\bibitem{holdom} B. Holdom, Phys. Rev. D {\bf 24}, 1441 (1981). 

\bibitem{walk2} B. Holdom, Phys. Lett. B {\bf 150}, 301 (1985). 

\bibitem{walk3} T. Appelquist, D. Karabali e L. C. R. Wijewardhana, Phys. Rev. Lett. {\bf 57}, 957 (1986).

\bibitem{walk4} T. Appelquist and L. C. R. Wijewardhana, Phys. Rev. D {\bf 36}, 568 (1987). 

\bibitem{walk5} T. Appelquist, M. Piai, and R. Shrock, Phys. Rev. D{\bf 69}, 015002 (2004).

\bibitem{walk6} T. Appelquist, M. Piai and R. Shrock, Phys. Lett. B {\bf 593} , 175 (2004).

\bibitem{walk7} T. Appelquist and R. Shrock,  Phys. Rev. Lett. {\bf 90}, 201801-1 (2003).

\bibitem{walk8} T. Appelquist and R. Shrock, Phys. Lett. B {\bf 548} , 204 (2002).

\bibitem{walk9} M. Kurachi, R. Shrock  and K. Yamawaki,  Phys. Rev. D {\bf 76}, 035003 (2007).

\bibitem{peskin1} M. E. Peskin and T. Takeuchi, Phys. Rev. Lett. {\bf 65}, 964 (1990).

\bibitem{peskin2} M. E. Peskin and T. Takeuchi, Phys. Rev. D {\bf 46}, 381 (1992).

\bibitem{sannino1}  F. Sannino and K. Tuominen, Phys.  Rev. D {\bf 71}, 051901 (2005). 

\bibitem{sannino2}  R. Foadi, M. T. Frandsen, T. A. Ryttov and F. Sannino, Phys. Rev. D {\bf 76}, 055005 (2007).

\bibitem{sannino3} T. A. Ryttov and F. Sannino, Phys. Rev. D {\bf 78}, 115010 (2008).

\bibitem{yama1} V. A. Miransky and K. Yamawaki, Mod. Phys. Lett. A {\bf 4}, 129 (1989).

\bibitem{yama2} K.-I. Kondo, H. Mino and K. Yamawaki, Phys. Rev. D{\bf 39}, 2430 (1989).

\bibitem{mira2} V. A. Miransky, T. Nonoyama and K. Yamawaki, Mod. Phys. Lett. A{\bf 4}, 1409 (1989).

\bibitem{yama3} T. Nonoyama, T. B. Suzuki and K. Yamawaki, Prog. Theor. Phys.{\bf 81}, 1238 (1989).

\bibitem{mira3} V. A. Miransky, M. Tanabashi and K. Yamawaki, Phys. Lett. B{\bf 221}, 177 (1989).

\bibitem{yama4} K.-I. Kondo, M. Tanabashi and K. Yamawaki, Mod. Phys. Lett. A{\bf 8}, 2859 (1993).

\bibitem{tak} T. Takeuchi, Phys. Rev. D {\bf 40}, 2697 (1989).

\bibitem{us1} A. Doff, A. A. Natale and P. S. Rodrigues da Silva, Phys. Rev. D {\bf 77}, 075012 (2008).

\bibitem{us2} A. Doff, A. A. Natale and P. S. Rodrigues da Silva, Phys. Rev. D {\bf 80}, 055005 (2009).

\bibitem{us4} A. Doff, E. G. S. Luna and A. A. Natale, Phys. Rev. D {\bf 88}, 055008 (2013).

\bibitem{twoscale} A. Doff and A. A. Natale, Phys. Lett. B {\bf 748}, 55 (2015).

\bibitem{us3} A. Doff and A. A. Natale, Int. J. Mod. Phys. A {\bf 31}, 1650024 (2016).

\bibitem{cg} A. Cohen and H. Georgi, Nucl. Phys. B {\bf 314}, 7 (1989).

\bibitem{us5} A. Doff and A. A. Natale, Phys. Lett. B {\bf 771}, 392 (2017). 

\bibitem{ardn} A. C. Aguilar ,  A. Doff and A. A. Natale,  hep-ph/1802.03206.

\bibitem{far} E. Farhi and L. Susskind, Phys. Rev. D {\bf 20}, 3404 (1979).

\bibitem{craig} Craig D. Robertz and Bruce H. J. McKellar,  Phys. Rev. D  {\bf 41}, 672 (1990).

\bibitem{mg1} J. M. Cornwall, Phys. Rev. D {\bf 26}, 1453 (1982).

\bibitem{mg2} A. C. Aguilar, D. Binosi and J. Papavassiliou, Phys. Rev. D {\bf 78}, 025010 (2008).

\bibitem{mg3} A. C. Aguilar and J. Papavassiliou, Phys. Rev. D {\bf 83}, 014013 (2011).

\bibitem{mg4} A. Doff, F. A. Machado and A. A. Natale, Annals of Physics {\bf 327}, 1030 (2012).




\end {thebibliography}
\end{document}